\documentclass[sigconf,review=False]{acmart}
\setcopyright{none}
\acmConference[ESEC/FSE 2020]{The 28th ACM Joint European Software Engineering Conference and Symposium on the Foundations of Software Engineering}{8 - 13 November, 2020}{Sacramento, California, United States}

\def\BibTeX{{\rm B\kern-.05em{\sc i\kern-.025em b}\kern-.08emT\kern-.1667em\lower.7ex\hbox{E}\kern-.125emX}}

\usepackage{graphicx}
\usepackage{array}
\usepackage{url}
\usepackage{fancybox}
\usepackage{multirow}
\usepackage{color}
\usepackage{colortbl}
\usepackage{balance}
\usepackage{fancyhdr}
\usepackage{subfig}
\usepackage{diagbox}
\usepackage{bbding}
\usepackage{placeins}
\usepackage{booktabs}
\usepackage{enumitem}
\usepackage{xcolor,lipsum}

\usepackage{textcomp}
\usepackage{algorithmic}
\usepackage{listings}
\usepackage{color}
\usepackage{commath}
\usepackage{comment}

\definecolor{light-gray}{rgb}{.906,  .902,  .902}

\usepackage{bm}

\begin{document}
% \nolinenumbers
% Title portion. Note the short title for running heads
\title{Code2Que: A Tool for Improving Question Titles from Mined Code Snippets in Stack Overflow}

\begin{abstract}
Stack Overflow is one of the most popular technical Q\&A sites used by software developers. Seeking help from Stack Overflow has become an essential part of software developers' daily work for solving programming-related questions.
Although the Stack Overflow community has provided quality assurance guidelines to help users write better questions, we observed that a significant number of questions submitted to Stack Overflow are of low quality. In this paper, we introduce a new web-based tool, {\sc Code2Que}, which can help developers in writing higher quality questions for a given code snippet. 
{\sc Code2Que} consists of two main stages: offline learning and online recommendation. 
In the offline learning phase, we first collect a set of good quality $\langle$\textit{code snippet, \textit{question}}$\rangle$ pairs as training samples. We then train our model on these training samples via a deep sequence-to-sequence approach,  enhanced with an \emph{attention} mechanism, a \emph{copy} mechanism and a \emph{coverage} mechanism.  
In the online recommendation phase, for a given code snippet, we use the offline trained model to generate question titles to assist less experienced developers in writing questions more effectively. 
At the same time, we embed the given code snippet into a vector and retrieve the related questions with similar problematic code snippets.
To evaluate {\sc Code2Que}, we first sampled 50 low quality $\langle$\textit{code snippet, \textit{question}}$\rangle$ pairs from the Python and Java datasets on Stack Overflow. Then we conducted a user study to evaluate the question titles generated by our approach as compared to human-written ones using three metrics: \textit{Clearness}, \textit{Fitness} and \textit{Willingness to Respond}. Our experimental results show that for a large number of low-quality questions in Stack Overflow, {\sc Code2Que} can improve the question titles in terms of \textit{Clearness}, \textit{Fitness} and \textit{Willingness} measures. 
{\sc Code2Que} can be accessed at \url{http://www.code2que.com}. A demo video of {\sc Code2Que} is at \url{https://youtu.be/orG--uXKnkU}.  
\end{abstract}

%
% The code below should be generated by the tool at
% http://dl.acm.org/ccs.cfm
% Please copy and paste the code instead of the example below.
%
% \begin{CCSXML}
% <ccs2012>
% <concept>
% <concept_id>10011007.10011074.10011111.10011113</concept_id>
% <concept_desc>Software and its engineering~Software evolution</concept_desc>
% <concept_significance>500</concept_significance>
% </concept>
% <concept>
% <concept_id>10011007.10011074.10011111.10011696</concept_id>
% <concept_desc>Software and its engineering~Maintaining software</concept_desc>
% <concept_significance>500</concept_significance>
% </concept>
% </ccs2012>
% \end{CCSXML}

% \ccsdesc[500]{Software and its engineering~Software evolution}
% \ccsdesc[500]{Software and its engineering~Maintaining software}
%
% End generated code
%
\author{Zhipeng Gao}
\affiliation{%
\institution{Monash University, Australia}
}
\author{Xin Xia}
\affiliation{%
\institution{Monash University, Australia}
}  
\author{David Lo}
\affiliation{%
\institution{Singapore Management University, Singapore}
}
\author{John Grundy}
\affiliation{%
\institution{Monash University, Australia}
}
\author{Yuan-Fang Li}
\affiliation{%
\institution{Monash University, Australia}
}

%\keywords{Smart Contract, Clone Detection, Bug Detection, Code Embedding}

% The default list of authors is too long for headers.
\renewcommand{\shortauthors}{Gao et al.}

\maketitle

\section{Introduction}
\label{sec:intro}
In recent years, Stack Overflow (SO) has become one of the most common ways that developers seek programming problem-related answers on the web. 
Millions of developers now use Stack Overflow 
to search for high-quality posts to solve their daily work problems. The success of Stack Overflow and community-based question and answer sites in general rely heavily on the will of community members to answer other's questions.
Intuitively, a well-phrased question is more likely to obtain attention from potential experts, thus increasing the likelihood of receiving useful help and support.
In contrast, poorly asked questions may discourage potential helpers and are less likely to receive useful answers, or indeed any answer at all.

Even though Stack Overflow has provided detailed guidelines to help community members post well-written questions, a large number of questions submitted to Stack Overflow are of low quality. These poorly asked questions are, more often than not, ambiguous, vague, and/or incomplete. 
It is thus very hard to attract potential experts to provide useful answers, which may discourage the askers and hinder the progress of knowledge sharing. 
Many prior works have investigated the issue of question quality on Stack Overflow~\cite{correa2014chaff, arora2015good, Trienes2019IdentifyingUQ}. 
Correa and Sureka~\cite{correa2014chaff} investigated closed questions on Stack Overflow, which suggests that a good question should contain enough code for others to reproduce the problem.
Arora et al.~\cite{arora2015good} proposed a novel method for improving the question quality prediction accuracy by making use of content extracted from previously asked similar questions in the forum. More recent work~\cite{Trienes2019IdentifyingUQ} studied approaches to identifying unclear questions in CQA websites.
However, all of these previous works focus on identifying the low-quality questions or how to increase the accuracy of the prediction, more in-depth research of improving the low-quality questions is still needed.

Based on our previous work~\cite{gao2020generating}, we present a web-based tool, named {\sc Code2Que}, that helps developers post higher-quality questions on Stack Overflow.
Developers can copy and paste their code snippets into our web application, then {\sc Code2Que} helps developers to improve their posts by generating question titles as well as recommending related questions.
The input to {\sc Code2Que} is a code snippet, which is regarded as an ordered sequence of code tokens by our tool.
{\sc Code2Que} consists of an offline learning phase and an online recommendation phase. 
The output of {\sc Code2Que} is two parts: (i) Generated Questions: {\sc Code2Que} will generate a high-quality question title for a given code snippet based on our deep sequence-to-sequence model. Developers can  utilize the generated questions for reformulating their posts. 
(ii) Retrieved Questions: {\sc Code2Que} first embeds the code snippet into a vector, then searches through our codebase to retrieve relevant questions with similar problematic code snippets. 
No prior work has provided practical tools to identify related questions via measuring similarity between code snippets. To the best of our knowledge, our work is the very first to investigate the possibility of automatically improving low-quality questions in Stack Overflow.

\begin{figure*}\vspace{-0.0cm}
\centerline{\includegraphics[width=0.90\textwidth]{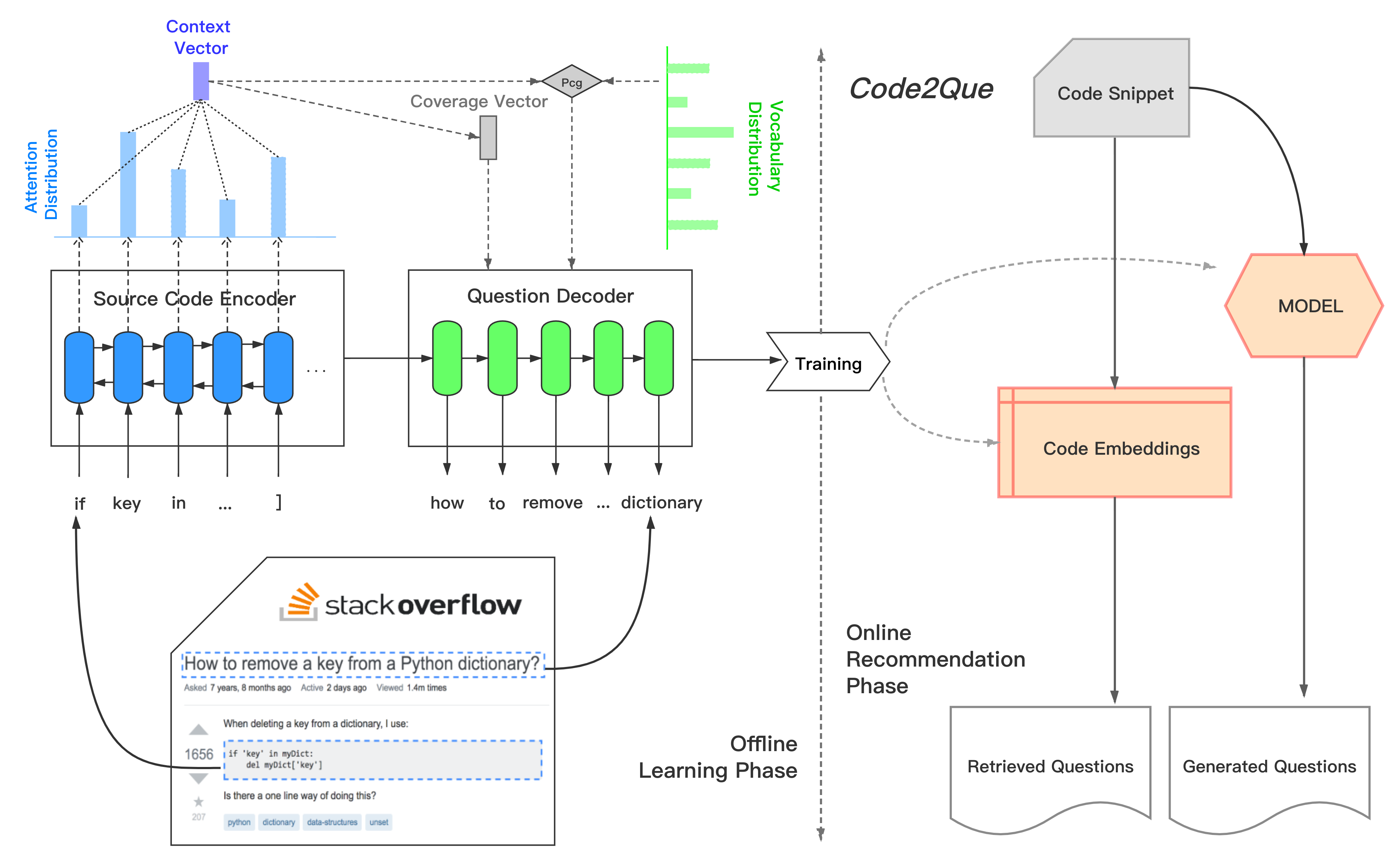}}
% \vspace*{-20pt}
\caption{Overview of Our Approach}
\label{fig:approach}
% \vspace{-0.2cm}
\end{figure*}

% To the best of our knowledge, this is the first that investigates the possibility of automatically improving low-quality questions in Stack Overflow. 
{\sc Code2Que} is able to benefit the following tasks: (i) Question Improvement. As many less experienced users lack the knowledge and terminology related to their problems or they have poor English writing skills, it is very hard, if not impossible, for them to always create clear and informative question titles. With the help of {\sc Code2Que}, developers can use the questions generated by our approach to reformulate their low-quality questions. Developers can also quickly review a list of related questions in Stack Overflow of similar problematic code snippets. They thus can gain a better understanding of the problem and revise their earlier poorly asked questions. (ii) Code Embeddings. Our approach can embed code snippets into a fixed dimensional vector space, and with this a variety of applications such as code search (e.g.~\cite{gu2018deep, cambronero2019deep}), clone detection (e.g.~\cite{li2017cclearner, gao2019teccd}), code summarization (e.g.~\cite{iyer2016summarizing, hu2018deep}), and API recommendation (e.g.~\cite{gu2016deep, reinhardt2018augmenting}) can benefit from the code embeddings used in our study.

The rest of the paper is organized as follows.
Section~\ref{sec:approach} presents the overall workflow of our {\sc Code2Que} approach and provides details of each step. 
Section~\ref{sec:implem} introduces the implementation details of {\sc Code2Que} and its key usage scenarios. 
Section~\ref{sec:eval} shows the experimental results of our evaluation.
Section~\ref{sec:con} summarises our work.

% \section{Related Work}\label{sec:related}
% \input{related}

\section{Approach}
\label{sec:approach}
\subsection{Overview}\label{AA}
Fig.~\ref{fig:approach} illustrates the overall framework of {\sc Code2Que}.
For a given code snippet, {\sc Code2Que} assists developers in writing high-quality questions by automatically generating question titles and retrieving the related questions in Stack Overflow. Our model contains two phases: offline learning and online recommendation.

In the offline learning phase, 
we first collect $\langle$\textit{code snippet, \textit{question}}$\rangle$ pairs from Stack Overflow posts. Since our goal is generating high-quality questions to help developers, we remove all the pairs in which the question score is less than 1. We train a deep sequence-to-sequence (seq2seq) model to map a code snippet directly to a high quality question title. Our offline learning model is divided into two components: a \textbf{Source-code Encoder} and a \textbf{Question Decoder}.
The source code snippet is transformed by Source-code Encoder into a vector  representation, with the help of an \textit{attention} mechanism ~\cite{bahdanau2014neural} to perform better content selection, a \textit{copy} mechanism ~\cite{gu2016incorporating} to handle the rare-words problem, as well as a \textit{coverage} mechanism~\cite{tu2016modeling} to avoid meaningless repetitions.
The vector representation of the code snippet is then read by a Question Decoder to generate the target question titles. 

In the online recommendation phase, for a given code snippet, the recommendation output is a Generated Question and a set of Retrieved Questions. 
The question title generated by the offline learning model can assist developers in writing high-quality questions that are more informative and clear. The retrieved similar questions can be used by developers to help them better understand their problems.

\subsection{Offline Learning}
\subsubsection{Source-code Encoder}
Our Source-code Encoder is a two-layer bidirectional LSTM network.
Tokens in the code snippet are fed sequentially into the Source-code Encoder, which generates a sequence of hidden states.
For example, given $x_{t}$ is the input source code token at time step $t$, 
the Source-code Encoder will produce the hidden states $\overrightarrow{\mathbf{h}_{t}}$ and $\overleftarrow{\mathbf{h}_{t}}$ at time step $t$ for the forward pass and backward pass respectively. 
The hidden states from the forward and backward pass of the last layer of the source-code encoder are concatenated to form a state $s$.

\subsubsection{Question Decoder}
Our Question Decoder is a single-layer LSTM network, initialized with the state $s$ produced by the Source-code Encoder.
During training, at each time step $t$, the Question Decoder takes as input the embedding vector $y_{t-1}$ of the previous word and the previous state $s_{t-1}$, and concatenates them to produce the input of the LSTM network. The output of the LSTM network is regarded as the decoder hidden state $s_t$.
The Question Decoder produces one symbol at a time and stops when the END symbol is emitted. 
The only change with the Question Decoder at testing time is that it uses output from the previous word, since there is no access to a ground truth then.

\subsubsection{Incorporating the Attention Mechanism}
One challenging task with the sequence-to-sequence model is dealing with the long sequence input. A solution was proposed by Bahdanau et al.~\cite{bahdanau2014neural}, in which they introduced a technique called ``\textit{attention}'' which significantly improved the performance of sequence-to-sequence models in machine translation systems. 
We incorporate the \textit{attention} mechanism in our work and model the attention~\cite{bahdanau2014neural} distribution over words in the source code snippets, 
which allows the model to focus on the most relevant parts of the input sequence as needed.

\subsubsection{Incorporating a Coverage Mechanism}
Repetition is another challenge for attentional sequence-to-sequence models, where meaninglessly repeated words can be generated during the decoding process. As shown in Figure~\ref{fig:gq_examples}, ``post'' has been repeated twice by the attention model (highlighted with yellow color in Figure~\ref{fig:gq_examples}). To address this problem, we incorporate a \textit{coverage} mechanism~\cite{tu2016modeling} to avoid meaningless repetitions. The \textit{coverage} mechanism quantitatively emphasizes the coverage of sentence words and thus avoids generating repetitive text while decoding. % \re{Incorporating \textit{coverage} mechanism can effectively eliminate such repetitions.} \yf{this is repetitive, can be removed.}

\subsubsection{Incorporating a Copy Mechanism}
\begin{figure}[t]
    \centerline{\includegraphics[width=0.44\textwidth]{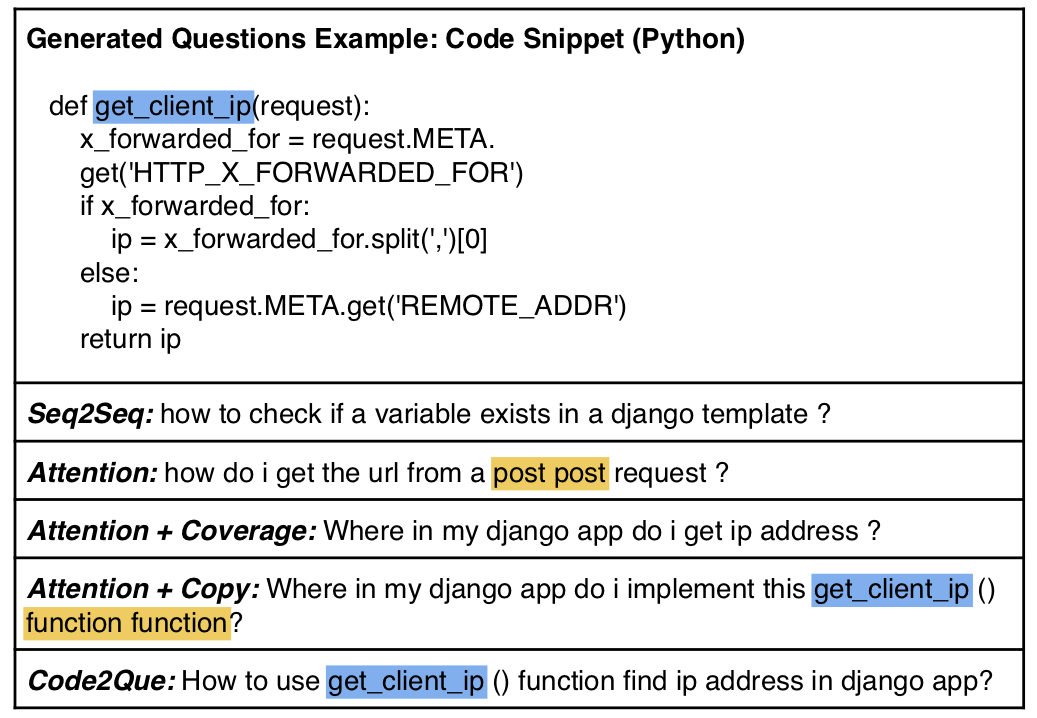}}
\vspace*{-10pt}
\caption{Example of Generated Questions}
\vspace*{-0.4cm}
\label{fig:gq_examples}
\end{figure}

% \todo{Add concrete example of how works??}
Generating question titles from code snippets is a non-trivial task because the code snippets usually contain tokens with very rare occurrences, such as the word \texttt{get\_client\_ip} (highlighted with a blue color) in Figure~\ref{fig:gq_examples}. It is very difficult, often impossible, for a decoder to generate such a word solely based on language modeling.
To address this challenge, we incorporate a \textit{copy} mechanism ~\cite{gu2016incorporating} which allows the model to copy tokens from the source code snippet to the target generated question title. To do this, we maintain a binary classifier $p_{cg}$ to determine whether to generate a word from the vocabulary or to copy the word directly from the input code snippet based on attention distribution. As shown in the last row in Figure~\ref{fig:gq_examples}, with the help of the \textit{copy} mechanism, the mohdel properly picks up the method name \texttt{get\_client\_ip} from the source code snippet and copies it into the generated question titles.

\subsection{Online Recommendation}
\subsubsection{Generated Questions.}
Once the offline learning model is trained, we do inference using a beam search on the pre-trained model.
For a given code snippet, the Source-code Encoder encodes it into a fixed-dimensional real-valued vector, then the Question Decoder reads the code embedding to infer the target question titles. The inference process stops when the model generates the END token which stands for the end of the sentence.
\subsubsection{Retrieved Questions.} To help developers better understand their problems, we retrieve other relevant questions in Stack Overflow according to the code snippet. 
After the offline training phase, each code snippet $s_{i}$ in the training corpus is mapped to a fixed-dimensional vector $c_{i}$ of real values. By stacking all the individual vectors together, we construct a source code snippet embedding matrix $C^{s \times d}$, where the first dimension $s$ is the total number of code snippets and the second dimension $d$ is the number of hidden states of the Source-code Encoder. After the developer submits his/her code snippet to our model, the code snippet is embedded into a vector by the Source-code Encoder, then we search through the code embedding matrix $C^{s \times d}$ to retrieve relevant questions with similar code snippets. % \yf{do you want to briefly say how similarity is calculated, e.g. cosine similarity, etc.?}

\section{Implementation Details}
\label{sec:implem}

We have implemented {\sc Code2Que} as a standalone web-based tool to assist developers in improving question titles in Stack Overflow.
The source code and data can be found in our Github repository\footnote{\url{https://github.com/beyondacm/Code2Que}}.

\smallskip
\noindent{\itshape\bfseries Data Collection}.
The data source of {\sc Code2Que} is from the Stack Overflow data dump of September 2019\footnote{\url{https://archive.org/details/stackexchange}}. We use the \textit{Python}, \textit{Java}, \textit{Javascript}, \textit{C\#} and \textit{SQL} tags to collect questions associated with the corresponding programming language. We extract the code snippet from each post's body by using $\langle$\textit{code}$\rangle$ tags, and then pair it with its question title if the question score is higher than 1. 
We have collected more than 1 million $\langle$\textit{code snippet}, \textit{question}$\rangle$ pairs from Stack Overflow for different programming languages.

\smallskip
\noindent{\itshape\bfseries Data Preprocessing}.
We preprocess our collected data according to the following steps: We first tokenize the code snippet and question title using the NLTK toolkit~\cite{bird2004nltk} (Step1). After that, to  avoid being overly context-specific, we use regular expressions to replace numbers and strings with special tokens ``NUMBER'' and ``STRING'' in code snippets (Step2). 
For question titles, we only keep the pairs if one of the interrogative keywords (e.g., ``how'', ``what'', ``why'', ``which'') appears in the question (Step3). Following that, we remove pairs where the code snippet and the question title are too long or too short. We set the token range from 16 tokens to 128 tokens for code snippets and from 4 tokens to 16 tokens for question titles respectively.  
The remaining $\langle$\textit{code snippet}, \textit{question}$\rangle$ pairs are added to our training corpus.

\begin{figure}[t]
    \centerline{\includegraphics[width=0.35\textwidth]{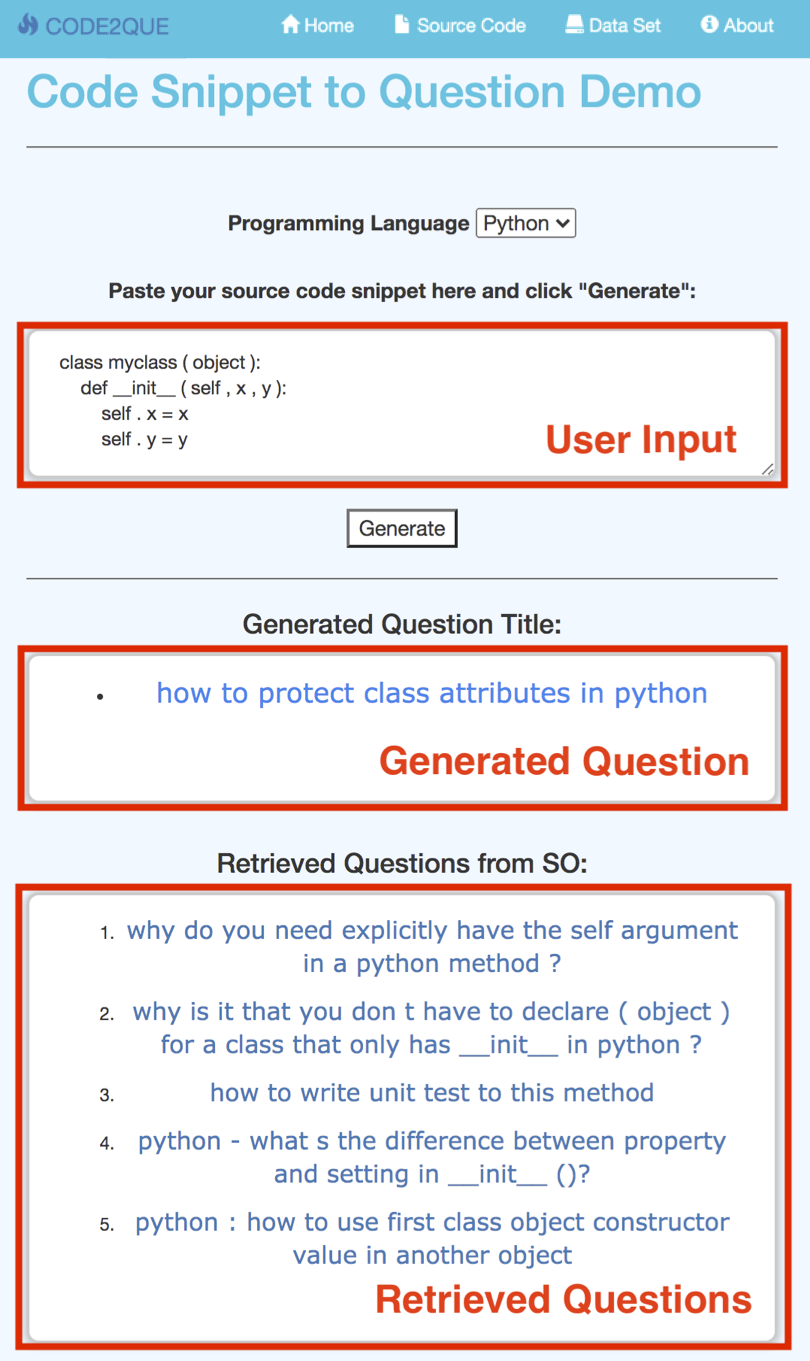}}
\vspace*{-10pt}
\caption{Homepage of Code2Que}
\vspace*{-0.5cm}
\label{fig:homepage}
\end{figure}

\smallskip
\noindent{\itshape\bfseries Backend Model}.
We use the framework OpenNMT~\cite{opennmt} to train our backend Model. 
We followed our previous experiment settings ~\cite{gao2020generating} for the training process.
The collected $\langle$\textit{question}, \textit{code snippet}$\rangle$ pairs are input into the workflow of our approach described in Section ~\ref{sec:approach}, and the output consists of the trained sequence-to-sequence model and the associated code embeddings which are used by {\sc Code2Que} as the backend model for online recommendation.

\smallskip
\noindent{\itshape\bfseries Frontend User Interface}.
%\subsection{Frontend User Interface}
Fig.~\ref{fig:homepage} shows the query page of {\sc Code2Que}.
{\sc Code2Que} provides an input box for developers to submit their source code snippet. After a developer submits his/her code snippet to the server, the code snippet is preprocessed and passed to our backend model. The outputs are organised into two separate result boxes for the generated question and retrieved questions respectively. For the generated question, {\sc Code2Que} generates a question title according to the code snippet.  For retrieved questions, {\sc Code2Que} returns the top-5 most similar questions in our code base together with the links to their Stack Overflow posts. 

\smallskip
\noindent{\itshape\bfseries System Optimization}.
To meet the efficiency requirement as an online web tool, we optimize the implementations of {\sc Code2Que}: 
(i) Considering that the offline learning model and code embeddings are frequently used in the online recommendation phase, we put them into cache to reduce redundant data loading. (ii) We created an index for the relevant questions in our database to speed up retrieval. (iii) We used locality sensitive hashing for fast nearest neighbour search in our large code embeddings data sets.

% \begin{figure}[t]
%     \centerline{\includegraphics[width=0.43\textwidth]{tool-bug-detection-print.pdf}}
% \vspace*{-10pt}
% \caption{Sample Results of Bug Detection}\vspace*{-0.6cm}
% \label{fig:tool-bug}
% \end{figure}

\section{Evaluation and User Study}
\label{sec:eval}
To evaluate whether {\sc Code2Que} can generate better question titles for low-quality questions in Stack Overflow, we performed a user study on our Python and Java datasets. We sampled 50 low-quality $\langle$\textit{question}, \textit{code snippet}$\rangle$ pairs, which have been marked as lacking clarity and/or need to be further improved upon. 
For each code snippet, we conducted a pairwise comparison between two question titles, the original title written by humans and the title generated by our {\sc Code2Que}. For each pairwise comparison, we asked 5 evaluators to decide which one is better in terms of three metrics: \emph{Clearness}, \emph{Fitness}, and \emph{Willingness}; tie was allowed. 
\emph{Clearness} measures whether a question title is expressed in a clear way.
\textit{Fitness} measures whether a question title is reasonable in logic with the provided code snippet. 
\textit{Willingness} measures whether a user is willing to respond to a specific question.

Evaluation results are summarized in Table~\ref{tab:manual_evaluation}. We can see that: (i) question titles generated by {\sc Code2Que}  outperform the low-quality question titles for all metrics. This demonstrates that our approach is able to produce clearer and/or more appropriate questions, indicating the ability of {\sc Code2Que} to improve low-quality Stack Overflow questions. (ii) {\sc Code2Que} question titles have better willingness scores. This shows that our question titles are more likely to elicit further interactions, and helpful to increase the likelihood of receiving answers. (iii) Not all of the poor quality question titles can be improved by our approach. Even though our question titles are not perfect, our {\sc Code2Que} is the first work on this direction. We release our tool and dataset to inspire further follow-up research.

\begin{table} \vspace{-0.0cm}
\caption{Human Evaluation (Code2Que vs. Human)}
\begin{center}
\vspace{-0.1cm}
\begin{footnotesize}
\begin{tabular}{|| c | c | c | c ||}
    \hline
    \ Python / Java & Win (\%) & Lose (\%) & Tie (\%) \\
    \hline\hline
    Clearness & 52.4 / 42.8 & 33.2 / 34.0 & 14.4 / 23.2 \\
    \hline
    Fitness   & 55.2 / 47.2 & 24.0 / 39.6 & 20.8 / 13.2 \\
    \hline
    Willingness  & 61.2 / 49.2 & 31.6 / 26.8 & 7.2 / 24.0 \\
    \hline
\end{tabular}
\end{footnotesize}
\label{tab:manual_evaluation}
\end{center}
\vspace{-0.5cm}
\end{table}

\section{Summary and Future Work}
\label{sec:con}
We demonstrate {\sc Code2Que}, a web-based tool for improving low-quality question titles in Stack Overflow. 
Developers copy and paste their code snippets into our web application. {\sc Code2Que} automatically generates question titles for the code snippets via a deep sequence-to-sequence model. Developers can utilize the generated questions for their posts. {\sc Code2Que} also searches for related questions with similar problematic code snippets to help developers better understand their problems. Human evaluation results show that for a large number of low-quality questions in Stack Overflow, {\sc Code2Que} can improve the quality of the question titles in terms of \emph{Clearness}, \emph{Fitness}, and \emph{Willingness}. In future work, we will design better models to generate question titles by considering more context information.

\balance
\bibliographystyle{ACM-Reference-Format}
\bibliography{samples}

%%% -*-BibTeX-*-
%%% Do NOT edit. File created by BibTeX with style
%%% ACM-Reference-Format-Journals [18-Jan-2012].

\begin{thebibliography}{17}

%%% ====================================================================
%%% NOTE TO THE USER: you can override these defaults by providing
%%% customized versions of any of these macros before the \bibliography
%%% command.  Each of them MUST provide its own final punctuation,
%%% except for \shownote{}, \showDOI{}, and \showURL{}.  The latter two
%%% do not use final punctuation, in order to avoid confusing it with
%%% the Web address.
%%%
%%% To suppress output of a particular field, define its macro to expand
%%% to an empty string, or better, \unskip, like this:
%%%
%%% \newcommand{\showDOI}[1]{\unskip}   % LaTeX syntax
%%%
%%% \def \showDOI #1{\unskip}           % plain TeX syntax
%%%
%%% ====================================================================

\ifx \showCODEN    \undefined \def \showCODEN     #1{\unskip}     \fi
\ifx \showDOI      \undefined \def \showDOI       #1{#1}\fi
\ifx \showISBNx    \undefined \def \showISBNx     #1{\unskip}     \fi
\ifx \showISBNxiii \undefined \def \showISBNxiii  #1{\unskip}     \fi
\ifx \showISSN     \undefined \def \showISSN      #1{\unskip}     \fi
\ifx \showLCCN     \undefined \def \showLCCN      #1{\unskip}     \fi
\ifx \shownote     \undefined \def \shownote      #1{#1}          \fi
\ifx \showarticletitle \undefined \def \showarticletitle #1{#1}   \fi
\ifx \showURL      \undefined \def \showURL       {\relax}        \fi
% The following commands are used for tagged output and should be
% invisible to TeX
\providecommand\bibfield[2]{#2}
\providecommand\bibinfo[2]{#2}
\providecommand\natexlab[1]{#1}
\providecommand\showeprint[2][]{arXiv:#2}

\bibitem[\protect\citeauthoryear{Arora, Ganguly, and Jones}{Arora
  et~al\mbox{.}}{2015}]%
        {arora2015good}
\bibfield{author}{\bibinfo{person}{Piyush Arora}, \bibinfo{person}{Debasis
  Ganguly}, {and} \bibinfo{person}{Gareth~JF Jones}.}
  \bibinfo{year}{2015}\natexlab{}.
\newblock \showarticletitle{The good, the bad and their kins: Identifying
  questions with negative scores in stackoverflow}. In
  \bibinfo{booktitle}{\emph{2015 IEEE/ACM International Conference on Advances
  in Social Networks Analysis and Mining (ASONAM)}}. IEEE,
  \bibinfo{pages}{1232--1239}.
\newblock


\bibitem[\protect\citeauthoryear{Bahdanau, Cho, and Bengio}{Bahdanau
  et~al\mbox{.}}{2014}]%
        {bahdanau2014neural}
\bibfield{author}{\bibinfo{person}{Dzmitry Bahdanau},
  \bibinfo{person}{Kyunghyun Cho}, {and} \bibinfo{person}{Yoshua Bengio}.}
  \bibinfo{year}{2014}\natexlab{}.
\newblock \showarticletitle{Neural machine translation by jointly learning to
  align and translate}.
\newblock \bibinfo{journal}{\emph{arXiv preprint arXiv:1409.0473}}
  (\bibinfo{year}{2014}).
\newblock


\bibitem[\protect\citeauthoryear{Bird and Loper}{Bird and Loper}{2004}]%
        {bird2004nltk}
\bibfield{author}{\bibinfo{person}{Steven Bird} {and} \bibinfo{person}{Edward
  Loper}.} \bibinfo{year}{2004}\natexlab{}.
\newblock \showarticletitle{NLTK: the natural language toolkit}. In
  \bibinfo{booktitle}{\emph{Proceedings of the ACL 2004 on Interactive poster
  and demonstration sessions}}. Association for Computational Linguistics,
  \bibinfo{pages}{31}.
\newblock


\bibitem[\protect\citeauthoryear{Cambronero, Li, Kim, Sen, and
  Chandra}{Cambronero et~al\mbox{.}}{2019}]%
        {cambronero2019deep}
\bibfield{author}{\bibinfo{person}{Jose Cambronero}, \bibinfo{person}{Hongyu
  Li}, \bibinfo{person}{Seohyun Kim}, \bibinfo{person}{Koushik Sen}, {and}
  \bibinfo{person}{Satish Chandra}.} \bibinfo{year}{2019}\natexlab{}.
\newblock \showarticletitle{When deep learning met code search}. In
  \bibinfo{booktitle}{\emph{Proceedings of the 2019 27th ACM Joint Meeting on
  European Software Engineering Conference and Symposium on the Foundations of
  Software Engineering}}. \bibinfo{pages}{964--974}.
\newblock


\bibitem[\protect\citeauthoryear{Correa and Sureka}{Correa and Sureka}{2014}]%
        {correa2014chaff}
\bibfield{author}{\bibinfo{person}{Denzil Correa} {and} \bibinfo{person}{Ashish
  Sureka}.} \bibinfo{year}{2014}\natexlab{}.
\newblock \showarticletitle{Chaff from the wheat: Characterization and modeling
  of deleted questions on stack overflow}. In
  \bibinfo{booktitle}{\emph{Proceedings of the 23rd international conference on
  World wide web}}. \bibinfo{pages}{631--642}.
\newblock


\bibitem[\protect\citeauthoryear{Gao, Wang, Liu, Yang, Sang, and Cai}{Gao
  et~al\mbox{.}}{[n.d.]}]%
        {gao2019teccd}
\bibfield{author}{\bibinfo{person}{Yi Gao}, \bibinfo{person}{Zan Wang},
  \bibinfo{person}{Shuang Liu}, \bibinfo{person}{Lin Yang},
  \bibinfo{person}{Wei Sang}, {and} \bibinfo{person}{Yuanfang Cai}.}
  \bibinfo{year}{[n.d.]}\natexlab{}.
\newblock \showarticletitle{TECCD: A Tree Embedding Approach for Code Clone
  Detection}. In \bibinfo{booktitle}{\emph{2019 IEEE International Conference
  on Software Maintenance and Evolution (ICSME)}}. IEEE,
  \bibinfo{pages}{145--156}.
\newblock


\bibitem[\protect\citeauthoryear{Gao, Xia, Grundy, Lo, and Li}{Gao
  et~al\mbox{.}}{2020}]%
        {gao2020generating}
\bibfield{author}{\bibinfo{person}{Zhipeng Gao}, \bibinfo{person}{Xin Xia},
  \bibinfo{person}{John Grundy}, \bibinfo{person}{David Lo}, {and}
  \bibinfo{person}{Yuan-Fang Li}.} \bibinfo{year}{2020}\natexlab{}.
\newblock \showarticletitle{Generating Question Titles for Stack Overflow from
  Mined Code Snippets}.
\newblock \bibinfo{journal}{\emph{arXiv preprint arXiv:2005.10157}}
  (\bibinfo{year}{2020}).
\newblock


\bibitem[\protect\citeauthoryear{Gu, Lu, Li, and Li}{Gu et~al\mbox{.}}{2016a}]%
        {gu2016incorporating}
\bibfield{author}{\bibinfo{person}{Jiatao Gu}, \bibinfo{person}{Zhengdong Lu},
  \bibinfo{person}{Hang Li}, {and} \bibinfo{person}{Victor~OK Li}.}
  \bibinfo{year}{2016}\natexlab{a}.
\newblock \showarticletitle{Incorporating copying mechanism in
  sequence-to-sequence learning}.
\newblock \bibinfo{journal}{\emph{arXiv preprint arXiv:1603.06393}}
  (\bibinfo{year}{2016}).
\newblock


\bibitem[\protect\citeauthoryear{Gu, Zhang, and Kim}{Gu et~al\mbox{.}}{2018}]%
        {gu2018deep}
\bibfield{author}{\bibinfo{person}{Xiaodong Gu}, \bibinfo{person}{Hongyu
  Zhang}, {and} \bibinfo{person}{Sunghun Kim}.}
  \bibinfo{year}{2018}\natexlab{}.
\newblock \showarticletitle{Deep code search}. In
  \bibinfo{booktitle}{\emph{2018 IEEE/ACM 40th International Conference on
  Software Engineering (ICSE)}}. IEEE, \bibinfo{pages}{933--944}.
\newblock


\bibitem[\protect\citeauthoryear{Gu, Zhang, Zhang, and Kim}{Gu
  et~al\mbox{.}}{2016b}]%
        {gu2016deep}
\bibfield{author}{\bibinfo{person}{Xiaodong Gu}, \bibinfo{person}{Hongyu
  Zhang}, \bibinfo{person}{Dongmei Zhang}, {and} \bibinfo{person}{Sunghun
  Kim}.} \bibinfo{year}{2016}\natexlab{b}.
\newblock \showarticletitle{Deep API learning}. In
  \bibinfo{booktitle}{\emph{Proceedings of the 2016 24th ACM SIGSOFT
  International Symposium on Foundations of Software Engineering}}.
  \bibinfo{pages}{631--642}.
\newblock


\bibitem[\protect\citeauthoryear{Hu, Li, Xia, Lo, and Jin}{Hu
  et~al\mbox{.}}{2018}]%
        {hu2018deep}
\bibfield{author}{\bibinfo{person}{Xing Hu}, \bibinfo{person}{Ge Li},
  \bibinfo{person}{Xin Xia}, \bibinfo{person}{David Lo}, {and}
  \bibinfo{person}{Zhi Jin}.} \bibinfo{year}{2018}\natexlab{}.
\newblock \showarticletitle{Deep code comment generation}. In
  \bibinfo{booktitle}{\emph{Proceedings of the 26th Conference on Program
  Comprehension}}. ACM, \bibinfo{pages}{200--210}.
\newblock


\bibitem[\protect\citeauthoryear{Iyer, Konstas, Cheung, and Zettlemoyer}{Iyer
  et~al\mbox{.}}{2016}]%
        {iyer2016summarizing}
\bibfield{author}{\bibinfo{person}{Srinivasan Iyer}, \bibinfo{person}{Ioannis
  Konstas}, \bibinfo{person}{Alvin Cheung}, {and} \bibinfo{person}{Luke
  Zettlemoyer}.} \bibinfo{year}{2016}\natexlab{}.
\newblock \showarticletitle{Summarizing source code using a neural attention
  model}. In \bibinfo{booktitle}{\emph{Proceedings of the 54th Annual Meeting
  of the Association for Computational Linguistics (Volume 1: Long Papers)}},
  Vol.~\bibinfo{volume}{1}. \bibinfo{pages}{2073--2083}.
\newblock


\bibitem[\protect\citeauthoryear{Klein, Kim, Deng, Senellart, and Rush}{Klein
  et~al\mbox{.}}{2017}]%
        {opennmt}
\bibfield{author}{\bibinfo{person}{Guillaume Klein}, \bibinfo{person}{Yoon
  Kim}, \bibinfo{person}{Yuntian Deng}, \bibinfo{person}{Jean Senellart}, {and}
  \bibinfo{person}{Alexander~M. Rush}.} \bibinfo{year}{2017}\natexlab{}.
\newblock \showarticletitle{Open{NMT}: Open-Source Toolkit for Neural Machine
  Translation}. In \bibinfo{booktitle}{\emph{Proc. ACL}}.
\newblock
\urldef\tempurl%
\url{https://doi.org/10.18653/v1/P17-4012}
\showDOI{\tempurl}


\bibitem[\protect\citeauthoryear{Li, Feng, Zhuang, Meng, and Ryder}{Li
  et~al\mbox{.}}{2017}]%
        {li2017cclearner}
\bibfield{author}{\bibinfo{person}{Liuqing Li}, \bibinfo{person}{He Feng},
  \bibinfo{person}{Wenjie Zhuang}, \bibinfo{person}{Na Meng}, {and}
  \bibinfo{person}{Barbara Ryder}.} \bibinfo{year}{2017}\natexlab{}.
\newblock \showarticletitle{Cclearner: A deep learning-based clone detection
  approach}. In \bibinfo{booktitle}{\emph{2017 IEEE International Conference on
  Software Maintenance and Evolution (ICSME)}}. IEEE,
  \bibinfo{pages}{249--260}.
\newblock


\bibitem[\protect\citeauthoryear{Reinhardt, Zhang, Mathur, and Kim}{Reinhardt
  et~al\mbox{.}}{2018}]%
        {reinhardt2018augmenting}
\bibfield{author}{\bibinfo{person}{Anastasia Reinhardt},
  \bibinfo{person}{Tianyi Zhang}, \bibinfo{person}{Mihir Mathur}, {and}
  \bibinfo{person}{Miryung Kim}.} \bibinfo{year}{2018}\natexlab{}.
\newblock \showarticletitle{Augmenting stack overflow with API usage patterns
  mined from GitHub}. In \bibinfo{booktitle}{\emph{Proceedings of the 2018 26th
  ACM Joint Meeting on European Software Engineering Conference and Symposium
  on the Foundations of Software Engineering}}. \bibinfo{pages}{880--883}.
\newblock


\bibitem[\protect\citeauthoryear{Trienes and Balog}{Trienes and Balog}{2019}]%
        {Trienes2019IdentifyingUQ}
\bibfield{author}{\bibinfo{person}{Jan Trienes} {and}
  \bibinfo{person}{Krisztian Balog}.} \bibinfo{year}{2019}\natexlab{}.
\newblock \showarticletitle{Identifying Unclear Questions in Community Question
  Answering Websites}. In \bibinfo{booktitle}{\emph{ECIR}}.
\newblock


\bibitem[\protect\citeauthoryear{Tu, Lu, Liu, Liu, and Li}{Tu
  et~al\mbox{.}}{2016}]%
        {tu2016modeling}
\bibfield{author}{\bibinfo{person}{Zhaopeng Tu}, \bibinfo{person}{Zhengdong
  Lu}, \bibinfo{person}{Yang Liu}, \bibinfo{person}{Xiaohua Liu}, {and}
  \bibinfo{person}{Hang Li}.} \bibinfo{year}{2016}\natexlab{}.
\newblock \showarticletitle{Modeling coverage for neural machine translation}.
\newblock \bibinfo{journal}{\emph{arXiv preprint arXiv:1601.04811}}
  (\bibinfo{year}{2016}).
\newblock


\end{thebibliography}

\end{document}